\RequirePackage[T1]{fontenc}
\documentclass[journal]{IEEEtran}
\usepackage{amsmath}
\usepackage{array}
\usepackage{graphicx}
\usepackage{cite}
\usepackage{float}
\usepackage{amssymb}
\usepackage{multirow}
\usepackage{booktabs}
\usepackage{xcolor}
\usepackage[breaklinks,hidelinks]{hyperref}

\begin{document}

\title{Covariance-Weighted Wavelength--Power Readout for a Cascaded Fiber Sensor}

\author{Xingnan Chen$^{\dagger}$, Riming Xu$^{\dagger}$, and Jiawen Zhang*
\thanks{Xingnan Chen and Riming Xu contributed equally to this work and share first authorship.}
\thanks{Xingnan Chen is with the College of Mechanical and Electrical Engineering, Hainan Vocational University of Science and Technology, Haikou, Hainan 571125, China.}
\thanks{Riming Xu is with the Physical Sciences and Engineering Division, King Abdullah University of Science and Technology, Thuwal 23955-6900, Saudi Arabia. Email: xu.rm@qq.com.}
\thanks{Jiawen Zhang is with the Department of Materials Science and Engineering, Stanford University, Stanford, California 94305, USA. Corresponding author: Jiawen Zhang. Email: zjiawen@stanford.edu.}}

\maketitle
\IEEEpeerreviewmaketitle

\begin{abstract}
Temperature, NaCl concentration, and liquid level jointly affect fiber-sensor spectra, complicating their simultaneous estimation. We fabricate a Sagnac-loop sensor by cascading three polarization-maintaining-fiber (PMF) sections: an untreated reference section, a thermally annealed section, and a locally arc-treated section. Separate wavelength and minimum-power regression models were trained on 27 joint parameter conditions, and their cross-fitted prediction errors determined bias corrections and parameter-specific fusion weights. On the same 18 evaluation spectra, we compared independent wavelength-matrix inversion, bias-corrected wavelength regression, and covariance-weighted wavelength--power fusion. Fusion reduced liquid-level root-mean-square error (RMSE) from 0.2448 to 0.1390~cm and concentration RMSE from 1.0594 to 0.9461 mass-fraction percentage points; paired bootstrap intervals excluded zero for both reductions. The temperature RMSE difference, 0.4035 versus 0.4284~$^\circ$C, lay within its interval. Adding power to the bias-corrected wavelength model further reduced liquid-level RMSE by 0.0088~cm, whereas the concentration power weight was zero. All 18 liquid-level errors were positive, and their mean accounted for 90\% of the fused mean-square error. A second-order response model fitted to the training spectra attributed most of this offset to curvature in the trough-wavelength responses, which linear estimators cannot represent and training-error bias correction cannot remove at intermediate conditions. Small wavelength differences between the training and evaluation measurements contributed most of the remaining temperature offset. A weaker spectral readout can therefore add a resolvable improvement when weighted by its prediction reliability, while response curvature is the main remaining error source.
\end{abstract}


\section{Introduction}
\IEEEPARstart{M}{onitoring} liquid processes often requires simultaneous measurement of temperature, composition, and liquid level. Compact optical-fiber probes support remote measurements in solution preparation, chemical processing, and liquid storage, with immunity to electromagnetic interference \cite{pevec2019multiparameter}. These quantities nevertheless affect the same spectrum: temperature changes both liquid refractive index (RI) and fiber response, composition changes RI, and immersion changes sensing length. Recovering all three therefore requires distinguishable spectral responses and an estimation method that uses their information effectively.

\subsection{Complementary Responses for Multiparameter Liquid Sensing}
Complementary responses can be obtained from different sensing regions, spectral features, or interference mechanisms. Tilted fiber Bragg gratings combine a core temperature reference with immersion-dependent cladding-mode amplitudes \cite{osuch2016liquid}. Surface-plasmon-resonance fiber tips combine resonance wavelength and depth responses for RI, temperature, and liquid-level measurement \cite{alonso2019spr}. Cascaded micro-tapered long-period gratings recover the same quantities from distinct resonance shifts \cite{yang2024cascadedtlpg}. Hybrid Sagnac structures separate RI from temperature or liquid level from strain by combining interference mechanisms \cite{zhao2017sagnac,chen2022liquid}. Functional and polarization-dependent responses provide further routes: laser-processed interferometric--plasmonic structures distinguish temperature and pH, while helical gratings in polarization-maintaining fiber (PMF) distinguish temperature and strain \cite{gong2024lowcrosstalk,jiang2022helical}.

Liquid sensing additionally requires distinguishing the fiber's thermal response from temperature-dependent changes in liquid RI. Differential white-light interferometry compensates the former while retaining the latter \cite{gouveia2013differential}, so fiber temperature compensation alone does not remove temperature effects from concentration estimates. Liquid-level sensitivity also varies with RI in a PMF--D-shaped multimode-fiber loop \cite{lu2021liquid}. Fourier-amplitude demodulation improves PMF Michelson liquid-level readout, yet the amplitude remains temperature dependent \cite{wang2022fftlevel}. These coupled effects motivate calibration across joint temperature, composition, and immersion conditions.

\subsection{PMF Spectral Design and Sensitivity Amplification}
PMF provides a tunable interference spectrum through birefringence, interaction length, and polarization coupling \cite{noda1986polarization}. Its response reflects both geometry and internal stress: drawing conditions affect residual stress, and stress-rod boron content changes PANDA-fiber beat length \cite{just2015stress,yu2025boron}. Hole-assisted designs combine geometric and stress effects, while enhanced geometric birefringence improves gyroscope thermal stability \cite{li2021holeassistance,li2023thermal}. Section arrangement and processing offer additional spectral control. Three-section loop filters tune fringe spacing and amplitude through polarization settings \cite{xu2024tunablecomb}; CO$_2$-laser annealing modifies stress-induced birefringence in PMF containing regenerated gratings \cite{lai2015annealing}. Directional arc treatment and subsequent thermal cycling have also formed and stabilized modal interference in polarization-maintaining photonic crystal fiber \cite{rong2014high}.

For liquid sensing, etched PMF loop mirrors reduce cladding thickness to obtain RI-sensitive wavelength shifts \cite{zhong2012absolute,wang2013liquid}. Tapered photonic crystal fibers and concatenated PMF tapers modify modal confinement for RI or salinity measurement \cite{pawar2016tapered,yang2021salinity}. Multitapered and macehead-shaped PMF sensors use wavelength or intensity responses to estimate RI and temperature \cite{liu2023multitapered,kumar2025macehead}. A PMF-assembled microfiber coupler uses wavelength primarily for RI and intensity primarily for temperature \cite{zhao2019microfiber}. These examples establish that trough positions and transmitted intensities can both carry parameter information, motivating their joint use in spectral readout.

Spectral displacement can also be amplified through the Vernier effect in cascaded interferometers \cite{liu2019vernier,zhang2022vernier}. Cascaded Sagnac loops, harmonic configurations, and opposite-sign thermal shifts increase temperature-dependent envelope displacement \cite{zhang2024cascaded,xie2022harmonic,tong2024enhanced}. Greater displacement, however, does not directly determine parameter-estimation accuracy. Zuo \emph{et al.} increased temperature sensitivity from $-1.134$ to $54.589$~nm/$^\circ$C using cascaded polarization-mode and Fabry--Perot interferometers. The reported detection limit changed from $3.98\times10^{-3}$ to $5.13\times10^{-3}$~$^\circ$C \cite{zuo2024ultrasensitive}. Reference gratings address fringe-order identification in high-sensitivity strain sensing and Vernier-enhanced cryogenic thermometry \cite{liu2023fsrrange,xu2026cryogenic}; Sagnac--Fabry--Perot configurations also extend to temperature--humidity measurement \cite{wu2025humidityvernier}. These developments highlight distinct roles for spectral sensitivity, unambiguous range, and the information used for parameter recovery.

\subsection{Parameter Estimation from Spectral Features}
Sensitivity-matrix inversion connects measured spectral shifts to parameter changes using calibrated response coefficients. Applications include two-parameter PMF sensing and three-trough strain, temperature, and torsion estimation \cite{liu2023multitapered,fu2024threeparameter}. A full-rank matrix permits unique linear estimates, while similarity between response directions determines how feature errors propagate through inversion \cite{pevec2019multiparameter}. Spectral peaks have accordingly been selected to minimize the condition number, improving strain and temperature estimates from a tilted fiber Bragg grating \cite{takeda2022tfbg}. The condition number, however, describes how random feature errors are amplified. Nonlinear wavelength responses can also produce large sensitivity-matrix errors, which has motivated machine-learning demodulation of fiber temperature--pressure and strain--temperature sensors \cite{chen2023seawaterml,wei2025cnnbilstm}. Calibration repeatability also matters: averaging sensitivities over repeated pressure cycles and updating the matrix reduced errors in a tapered-fiber temperature--pressure sensor \cite{lu2022temperaturepressure}.

Our previous cascaded single-mode-fiber--multimode-fiber--long-period-grating sensor established three-trough sensitivity responses to temperature, strain, and RI \cite{xu2026multitrough}. Its coupled experiments demonstrated temperature-compensated concentration sensing and temperature--strain decoupling, establishing the feasibility of multi-trough responses and dual-parameter discrimination. Building on this approach, the present study examines simultaneous estimation when temperature, NaCl concentration, and liquid level vary jointly. We further use the minimum transmitted power at each trough alongside its wavelength.

Feature choice links the optical response to the estimation task. Wavelength, trough depth, envelope position, and Fourier amplitude capture different spectral properties \cite{wang2022fftlevel}, while interrogation hardware balances resolution, speed, complexity, and stability \cite{dey2025interrogation}. Full-spectrum convolutional networks extend beyond tracked extrema, estimating salinity from multimode Mach--Zehnder spectra and pressure from Fabry--Perot spectra \cite{lv2024dcnn,hou2026cnn}. Grating-enhanced Mach--Zehnder spectra have also supported three-output networks trained on separately varied parameter series \cite{li2022gratingcnn}. Related studies examine Raman regression with omitted chemical components \cite{boateng2022raman} and interpretable networks for NaCl-film thickness, concentration, and temperature \cite{sun2026development}. Together with reviews of fiber-sensor learning \cite{cao2025artificial}, these approaches motivate matching the spectral representation and model to available calibration conditions.

Estimator fusion uses prediction reliability to determine how different spectral readouts contribute to parameter estimates. Combining estimates according to their prediction errors has a long history \cite{bates1969combination}. Covariance-based weighting describes both individual error magnitudes and shared errors \cite{bakr2018covariance}. For wavelength and minimum-power features extracted from the same spectrum, paired training prediction errors provide the statistics used to determine combination weights. Nonnegative weights assigned separately for each parameter can give a less accurate auxiliary estimate a small contribution or exclude it when the estimated combined risk favors a single readout. This motivates evaluating whether imperfect power information adds predictive value beyond trough wavelengths.

Different evaluation designs answer different questions about the accuracy of parameter predictions. Repeated measurements at calibrated conditions test the stability of a fixed estimator. Withholding complete parameter levels tests predictions at levels absent from training. Varying parameters separately during calibration also provides different information from measuring their joint combinations. In a tilted-grating Sagnac sensor for salinity and temperature, for example, the temperature error was larger when both parameters changed simultaneously than in the test set used to evaluate the trained network \cite{lin2024tfbgsagnac}. Related observations must be assigned to training and evaluation sets appropriately to avoid overstating performance \cite{kapoor2023leakage}. For estimator fusion, input scaling, bias correction, and prediction-error statistics must use training data alone. All methods should then be compared on the same evaluation spectra and reference values. The calibration data used to fit each method must also be specified. A sensitivity matrix calibrated by varying each parameter separately uses different training information from regressors fitted to joint parameter combinations. Comparing the fused estimates with the wavelength-only estimates helps distinguish improvements from joint calibration and bias correction from those obtained by adding power.

\subsection{Motivation and Scope of This Work}
Here, we fabricate a Sagnac-loop sensor with three cascaded PMF sections: an untreated reference section, a thermally annealed section, and a locally arc-treated section. Three tracked troughs provide paired wavelength and minimum-transmitted-power features for simultaneous estimation under joint temperature, NaCl-concentration, and liquid-level variation. Separate wavelength and power regression models estimate all three parameters. Their paired cross-fitted training errors determine regularized combination weights for each output, allowing the contribution of each readout to follow its predictive reliability.

The regression models use 27 joint calibration conditions, and training statistics determine fixed bias corrections and combination weights. We compare independent wavelength-matrix inversion, bias-corrected wavelength regression, and wavelength--power fusion on the same 18 evaluation spectra. Single-parameter wavelength sensitivities determine the matrix baseline. The wavelength-only comparison uses the same jointly calibrated wavelength model and bias correction as fusion, isolating the incremental contribution of power. Paired bootstrap intervals indicate which RMSE differences the 18 spectra resolve. We also separate each method's errors into a common offset and scatter about it. A second-order forward model fitted to the training spectra then distinguishes offsets caused by curvature in the spectral responses from those caused by spectral differences between the training and evaluation measurements. The study thus tests whether a weaker auxiliary readout can improve three-parameter estimation beyond joint wavelength calibration and bias correction, and identifies the error sources that remain.

\section{Spectral Data and Estimation Methods}
\subsection{Sensor preparation and optical readout}

The sensor consists of three cascaded PANDA polarization-maintaining-fiber (PMF) sections (YOFC, China) in a Sagnac loop, each with a cladding diameter of 125~$\mu$m. PMF-1 remains untreated as a reference, PMF-2 undergoes thermal annealing, and PMF-3 receives local arc treatment. PMF-1, PMF-2, and PMF-3 are each 3~cm long. The annealed region of PMF-2 is 3~cm long, and the arc-treated region of PMF-3 is 1~cm long. Before treatment, the polymer coating is removed from the processing regions of PMF-2 and PMF-3. These regions are then cleaned with absolute ethanol and dried with nitrogen.

For thermal annealing, PMF-2 is mounted straight on a quartz support with minimal axial tension and placed in a box furnace (Thermo Scientific Lindberg/Blue M BF51794C) under dry nitrogen. The fiber is heated from room temperature at 4~$^\circ$C/min to 430~$^\circ$C and held for 60~min. It is then allowed to cool naturally with the furnace closed and removed only after the temperature falls below 80~$^\circ$C.

Fusion splicing and local arc treatment use a Fujikura FSM-100P polarization-maintaining fusion splicer. For local arc treatment, PMF-3 is mounted with minimal tension on the discharge platform and connected to the complete Sagnac loop, with an optical spectrum analyzer (OSA) monitoring the spectrum. Arc discharges are applied at three positions in the central region of PMF-3, spaced 5~mm apart. The arc-power setting is $-20$~bit relative to the standard setting. Each discharge lasts 300~ms, with six discharges applied in total. Spectral monitoring checks for a stable, reproducible additional beat component. Microscopic inspection checks that the treated fiber has no visible collapse or damage.

Adjacent PMF sections are fusion-spliced with a polarization-axis rotation angle of $45^\circ$ to assemble the cascaded loop. The sensing section is installed vertically, so that its immersed length changes with the liquid level.

The optical readout uses a 1550-nm superluminescent diode (SLED) broadband source (Beijing Conquer Photonics) and a 1550-nm $2\times2$ single-mode fused 3-dB coupler (Thorlabs). An AQ6317B OSA records each spectrum in a single scan over 1400--1674~nm at 0.2-nm sampling intervals, with the resolution bandwidth set to 0.1~nm. No power normalization is applied to the acquired spectra; the OSA readings in dBm are used directly. Supplementary Note~S1 gives further preparation and acquisition details.

NaCl (Thermo Scientific, ACS reagent grade, purity $\geq99.0\%$) was weighed using a Fisherbrand FPRS423 precision balance. The salt was dissolved in Milli-Q water, and the solution was mixed thoroughly. A thermostatic water bath was used to control the solution temperature. After the measurement conditions stabilized, we waited an additional period before acquiring the spectra.

Liquid level is controlled by adjusting the NaCl solution volume in a large, open basin. The basin provides sufficient space for the sensing section, with the fiber leads at both ends extending outside the container. A millimetre-graduated ruler attached to the side of the basin is used to set the liquid level. The levels are set to 3.0, 3.5, 4.0, 4.5, 5.0, 5.5, and 6.0~cm by adding or withdrawing solution with a syringe. The ruler reading at each setting serves as the reference liquid level.

\begin{figure*}[t]
\centering
\includegraphics[width=\textwidth]{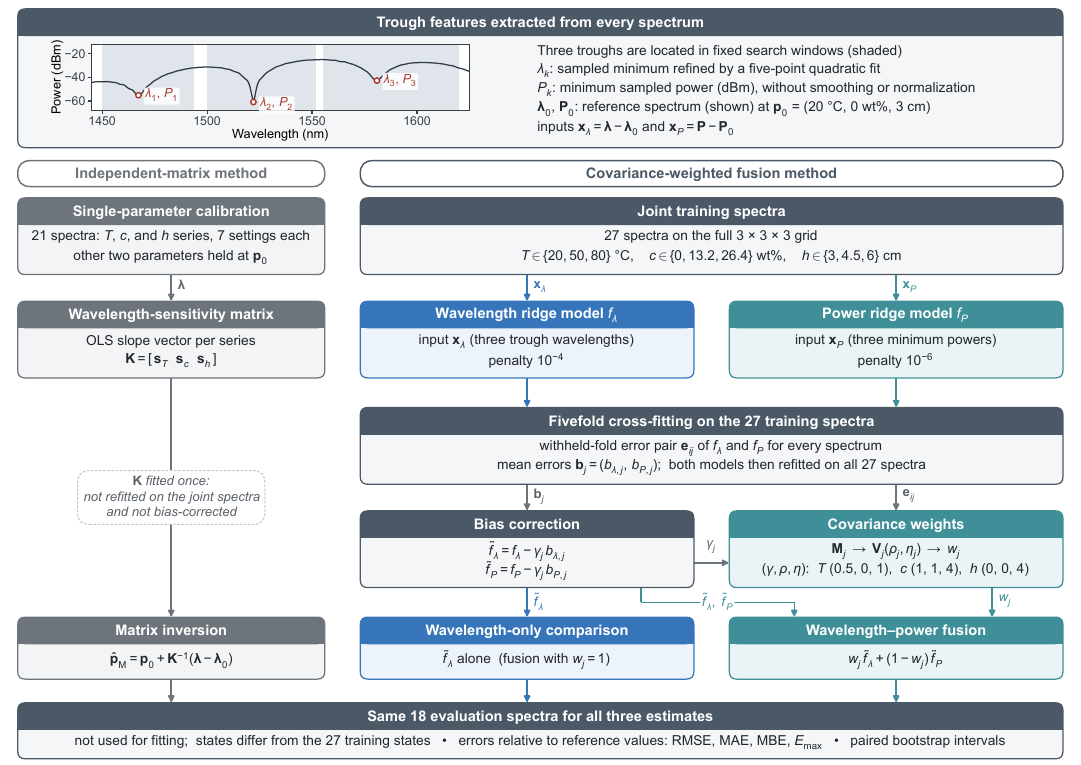}
\caption{Estimation workflow for temperature, NaCl concentration, and liquid level from measured spectra. The top panel shows the measured reference spectrum at $\mathbf p_0$, the three trough search windows, and the extracted trough features. Single-parameter calibration spectra determine the wavelength-sensitivity matrix used for matrix inversion. The joint training spectra fit separate wavelength and minimum-power regression models. Cross-fitted prediction errors determine the bias corrections and weights used to combine their estimates, with $(\gamma,\rho,\eta)$ fixed for each parameter. Here, $\tilde f_\lambda$ and $\tilde f_P$ denote the bias-corrected wavelength and power estimates. The bias-corrected wavelength estimate provides the wavelength-only comparison. All three estimates are evaluated on the same 18 spectra.}
\label{fig:cov_workflow}
\end{figure*}

We estimate temperature, NaCl concentration, and liquid level from the measured interference spectra. For each of three spectral troughs, we extract the wavelength and the minimum sampled power. The independent-matrix method estimates all three parameters from the wavelength shifts using a sensitivity matrix obtained from single-parameter calibrations. The fusion method uses two regression models: one receives the three wavelength features, and the other receives the three minimum-power features. Each model estimates temperature, concentration, and liquid level from its inputs. The two estimates of each parameter are then combined using weights calculated from their cross-fitted prediction errors. Figure~\ref{fig:cov_workflow} summarizes the steps in each estimation method.

The three parameters are written as $\mathbf p=(T,c,h)^{\mathsf T}$. Here, $T$ is temperature, $c$ is NaCl mass fraction in wt\%, and $h$ is vertical liquid level. We use the reference state $\mathbf p_0$ and divide parameter changes by the ranges in $\mathbf R$:
\begin{equation}
\begin{aligned}
\mathbf p_0&=(20~^\circ\mathrm C,0~\mathrm{wt}\%,3~\mathrm{cm})^{\mathsf T},\\
\mathbf R&=\operatorname{diag}(60~^\circ\mathrm C,26.4~\mathrm{pp},3~\mathrm{cm}).
\end{aligned}
\label{eq:cov_ranges}
\end{equation}
NaCl concentration ranges from 0 to 26.4~wt\%, so its normalization range is $26.4-0=26.4$ percentage points (pp).

\subsection{Spectral measurements, training, and evaluation}

We measured 234 spectra to characterize the sensor response and compare the estimation methods. These comprised 21 single-parameter calibration spectra, 30 repeated scans, 45 forward/return-cycle spectra, 93 fixed-condition stability spectra, and 45 joint-parameter spectra. Each spectrum was acquired in one scan, with 1371 wavelength samples from 1400 to 1674~nm at 0.2-nm intervals.

Single-parameter calibration comprised seven temperature measurements, seven NaCl-concentration measurements, and seven liquid-level measurements. Temperature was set to 20, 30, 40, 50, 60, 70, and 80~$^\circ$C, with concentration fixed at 0~wt\% and liquid level at 3~cm. Concentration was set to 0, 4.4, 8.8, 13.2, 17.6, 22, and 26.4~wt\%, with temperature fixed at 20~$^\circ$C and liquid level at 3~cm. Liquid level was set to 3, 3.5, 4, 4.5, 5, 5.5, and 6~cm, with temperature fixed at 20~$^\circ$C and concentration at 0~wt\%. These measurements determined how trough wavelength and minimum power changed with each parameter. Figure~\ref{fig:cov_calibration} shows the calibration spectra and their fitted responses.

Repeatability measurements comprised ten scans at a temperature of 50~$^\circ$C, ten scans at a NaCl concentration of 13.2~wt\%, and ten scans at a liquid level of 4.5~cm. The ten scans in each series were acquired at the same settings. Variation in trough wavelength and minimum power within each series characterized the consistency of repeated spectral acquisition.

For each parameter, five low--high--low cycles were measured, with one spectrum acquired at each of the three stages. The temperature sequence was $20\rightarrow80\rightarrow20~^\circ$C, the concentration sequence was $0\rightarrow26.4\rightarrow0$~wt\%, and the liquid-level sequence was $3\rightarrow6\rightarrow3$~cm. This gave 15 spectra per parameter and 45 spectra in total. The return spectrum in each cycle was compared with its starting spectrum to examine recovery after the parameter returned to its initial setting.

Stability was measured at three fixed conditions, $(T,c,h)=(50,0,3)$, $(20,13.2,3)$, and $(20,0,4.5)$, with units of $^\circ$C, wt\%, and cm, respectively. At each condition, spectra were collected every 4~min over 120~min. Including the initial measurement, each series contained 31 spectra, giving 93 spectra in total. Temperature, concentration, and liquid level remained fixed within each series to assess spectral variation over time.

The 27 joint training spectra covered every combination of these temperature, concentration, and liquid-level settings:
\begin{equation}
\begin{aligned}
T&\in\{20,50,80\}~^\circ\mathrm C,\\
c&\in\{0,13.2,26.4\}~\mathrm{wt}\%,\\
h&\in\{3,4.5,6\}~\mathrm{cm}.
\end{aligned}
\label{eq:cov_grid}
\end{equation}
We measured one spectrum at each of the 27 parameter combinations. These spectra trained the wavelength and minimum-power regression models and provided their prediction errors through cross-fitting.

We applied matrix inversion, bias-corrected wavelength regression, and fusion to the same 18 evaluation spectra and compared each estimate with the corresponding reference temperature, concentration, or liquid level. These spectra were measured within the calibration ranges at combinations different from the 27 training states. The same probe was used for both sets and was remounted between the joint training and evaluation measurements. Supplementary Note~S2 lists the reference values for all 18 spectra. None of these spectra or reference values was used to fit the regression models or estimate their cross-fitted errors. Repeated scans, cycle measurements, and stability measurements were analyzed separately from the training and evaluation spectra.

\subsection{Independent matrix calibration and inversion}

The matrix method uses the wavelengths of three spectral troughs to estimate temperature, concentration, and liquid level. We also extract the minimum power of each trough for the fusion method described in the next subsection.

We search for the three troughs within 1450--1494, 1500--1552, and 1555--1620~nm. The lowest sampled power within each window defines the trough power $P_i$ in dBm. A quadratic fit to five samples centered on that minimum refines the trough wavelength $\lambda_i$. We accept the fitted wavelength only if the curvature is positive and the shift from the sampled minimum is at most 0.4~nm. Otherwise, we retain the wavelength of the sampled minimum, without further refinement. The spectra are used as acquired, with no additional smoothing. The power feature remains the measured sample minimum; we do not use the fitted minimum or subtract the power at a trough shoulder. Power differences between values expressed in dBm are reported in dB.

For each single-parameter calibration series, we fit the extracted trough wavelengths against the varied parameter using ordinary least squares. The three seven-setting series give the wavelength-slope vectors $\mathbf s_T$, $\mathbf s_c$, and $\mathbf s_h$. These form the three columns of the wavelength-sensitivity matrix:
\begin{equation}
\mathbf K=
\begin{bmatrix}
\mathbf s_T & \mathbf s_c & \mathbf s_h
\end{bmatrix},
\label{eq:cov_matrix}
\end{equation}
The column units are nm/$^\circ$C, nm/pp, and nm/cm for temperature, concentration, and liquid level, respectively. Supplementary Note~S1 reports the wavelength and power slopes and their fit statistics. A separate intercept is fitted for each of the three calibration series. The reference wavelength and power vectors, $\boldsymbol\lambda_0$ and $\mathbf P_0$, come from the spectrum measured at $T=20~^\circ$C, $c=0$~wt\%, and $h=3$~cm, corresponding to $\mathbf p_0$. Matrix inversion estimates the three parameters as
\begin{equation}
\widehat{\mathbf p}_{\mathrm M}
=\mathbf p_0+\mathbf K^{-1}(\boldsymbol\lambda-\boldsymbol\lambda_0).
\label{eq:cov_inverse}
\end{equation}
The matrix is fitted once from the single-parameter calibration spectra. It remains unchanged when the 27 joint spectra are used to train the fusion method.

We normalize each matrix column as $\widetilde{\mathbf K}_{:j}=\mathbf K_{:j}/\|\mathbf K_{:j}\|_2$ to compare response directions independently of parameter units. The condition number $\kappa_2(\widetilde{\mathbf K})$ quantifies how well these directions can be distinguished. With the calibration matrix and reference spectrum fixed, additive wavelength-error covariance $\boldsymbol\Sigma_\lambda$ produces parameter-error covariance $\mathbf K^{-1}\boldsymbol\Sigma_\lambda\mathbf K^{-\mathsf T}$.

\subsection{Covariance-weighted wavelength--power fusion}

\subsubsection{Regression models for wavelength and power features}

The fusion method uses separate regression models for the three wavelength shifts and three minimum-power changes. Their inputs are $\mathbf x_\lambda=\boldsymbol\lambda-\boldsymbol\lambda_0$ and $\mathbf x_P=\mathbf P-\mathbf P_0$, respectively. Both models predict the normalized parameters $\mathbf y=\mathbf R^{-1}(\mathbf p-\mathbf p_0)$. Each model standardizes its inputs using training means and population standard deviations. Constant features are assigned a scale of one during standardization. Ridge regression then fits each model, with an unpenalized intercept. The penalties are fixed at $10^{-4}$ for wavelength and $10^{-6}$ for power. Each fit minimizes the sum of squared prediction errors plus a penalty, using normalized targets and standardized inputs. The penalty equals the fixed penalty coefficient times the sum of squared regression coefficients. Input standardization and regression fitting use only the spectra in the current training partition. NumPy implements the fit by centering standardized inputs and targets and solving the penalized normal equations.

Fivefold cross-fitting splits the current training set into five folds. Each model is fitted on four folds and predicts the remaining fold, repeating until every training spectrum has a withheld prediction. For parameter $j$, the errors from the two models and their mean are
\begin{equation}
\mathbf e_{ij}=
\begin{bmatrix}
f_\lambda^{(-i)}(\mathbf x_{\lambda,i})_j-y_{ij}\\
f_P^{(-i)}(\mathbf x_{P,i})_j-y_{ij}
\end{bmatrix},
\qquad
\mathbf b_j=\frac{1}{n}\sum_{i=1}^{n}\mathbf e_{ij}.
\label{eq:cov_errors}
\end{equation}
Here, $(-i)$ denotes a model fitted without the fold containing spectrum $i$. Each error pair contains the wavelength-model and power-model errors for the same spectrum. Fusion uses statistics of these prediction errors, expressed in normalized parameter units. It does not use covariance between the raw wavelength and power inputs.

\subsubsection{Covariance-weighted combination of the regression estimates}

We subtract a fraction $\gamma_j$ of the mean prediction error for parameter $j$. The second-moment matrix of the remaining errors is
\begin{equation}
\mathbf M_j=\frac{1}{n}\sum_i
(\mathbf e_{ij}-\gamma_j\mathbf b_j)
(\mathbf e_{ij}-\gamma_j\mathbf b_j)^{\mathsf T}.
\label{eq:cov_moment}
\end{equation}
Writing the centered error covariance with divisor $n$ as $\mathbf C_j$ gives $\mathbf M_j=\mathbf C_j+(1-\gamma_j)^2\mathbf b_j\mathbf b_j^{\mathsf T}$. The matrix therefore includes both error covariance and any bias left after correction. We then adjust the matrix using $\rho_j$ and $\eta_j$:
\begin{equation}
\begin{aligned}
\mathbf V_j&=\mathbf D_j
\left[\rho_j\mathbf M_j+(1-\rho_j)\operatorname{diag}(\mathbf M_j)\right]\mathbf D_j,\\
\mathbf D_j&=\operatorname{diag}(1,\sqrt{\eta_j}),
\end{aligned}
\label{eq:cov_risk}
\end{equation}
Here, $\operatorname{diag}(\mathbf M_j)$ retains only the diagonal entries of $\mathbf M_j$. The factor $\rho_j$ scales the off-diagonal entries linking errors from the two models. The factor $\eta_j$ multiplies the power-model diagonal entry and scales both off-diagonal entries by $\sqrt{\eta_j}$. These adjustments control how strongly the combination depends on error relationships and power predictions estimated from the small training set.

The two estimates receive weights $w_j$ and $1-w_j$. Minimizing their combined estimated squared error gives the wavelength-model weight \cite{bates1969combination,bakr2018covariance}
\begin{equation}
w_j=\operatorname{clip}_{[0,1]}
\left(\frac{V_{PP,j}-V_{\lambda P,j}}
{V_{\lambda\lambda,j}+V_{PP,j}-2V_{\lambda P,j}}\right).
\label{eq:cov_weight}
\end{equation}
If the denominator in normalized target units is at most $10^{-12}$, we use equal weights, $w_j=1/2$. Clipping $w_j$ to $[0,1]$ ensures that both model weights remain nonnegative. Temperature, concentration, and liquid-level estimates themselves are not clipped to their calibration ranges. The combined estimate is
\begin{equation}
\begin{aligned}
\widehat y_{\mathrm F,j}
={}&w_j\left[f_\lambda(\mathbf x_\lambda)_j-\gamma_j b_{\lambda,j}\right]\\
&+(1-w_j)\left[f_P(\mathbf x_P)_j-\gamma_j b_{P,j}\right],\\
\widehat{\mathbf p}_{\mathrm F}&=\mathbf p_0+\mathbf R\widehat{\mathbf y}_{\mathrm F}.
\end{aligned}
\label{eq:cov_fusion}
\end{equation}
The fixed settings are $(\gamma,\rho,\eta)=(0.5,0,1)$ for temperature, $(1,1,4)$ for concentration, and $(0,0,4)$ for liquid level. Fivefold cross-fitting on the 27 training spectra gives the mean errors, adjusted error matrices, and weights for these settings. Both regression models are then fitted to all 27 training spectra. The fitted models, bias corrections, and weights remain fixed during evaluation. Each spectrum is processed separately, without using earlier predictions or updating covariance over time.

\subsection{Cross-validation and comparison of estimation errors}

We assess the fixed fusion settings with four cross-validation schemes using only the 27 training spectra. The first pools predictions from shuffled fivefold validation. The other three withhold all nine states at the middle temperature, concentration, or liquid level, respectively. Each of these three schemes trains on the remaining 18 states and predicts the nine withheld states. For every outer training set, an inner fivefold cross-fit estimates the mean errors and error matrices used for fusion. The withheld outer states are excluded from regression fitting and estimation of these quantities. The independent matrix remains fixed from the single-parameter calibrations throughout these comparisons.

For each scheme, we report the fusion RMSE divided by the matrix-inversion RMSE. The four schemes reuse spectra and therefore do not constitute four independent datasets. After cross-validation, the final fusion model is fitted using all 27 training spectra as described above.

The cross-validation splits use fixed random seeds for reproducibility. Supplementary Note~S2 gives the fold assignments, software versions, and implementation details.

Our main comparison applies matrix inversion, bias-corrected wavelength regression, and wavelength--power fusion to the same 18 evaluation spectra. The wavelength-only comparison uses the same fitted wavelength model and bias correction as fusion, with no power contribution. Supplementary Note~S3 also reports the uncorrected wavelength estimates. For each parameter, we subtract the corresponding reference value from each estimate to calculate the error. All 18 spectra are retained, including those producing large errors, and no predictions are clipped.

For errors $\epsilon_i=\widehat p_i-p_i$, the reported statistics are
\begin{equation}
\begin{aligned}
\mathrm{RMSE}&=\sqrt{N^{-1}\sum_i\epsilon_i^2},&
\mathrm{MAE}&=N^{-1}\sum_i|\epsilon_i|,\\
\mathrm{MBE}&=N^{-1}\sum_i\epsilon_i,&
E_{\max}&=\max_i|\epsilon_i|.
\end{aligned}
\label{eq:cov_metrics}
\end{equation}
Concentration errors are reported in percentage points (pp), consistent with the normalization range defined above. These results describe estimation performance at the 18 historical evaluation conditions. Prediction RMSE alone does not determine the sensor's measurement resolution. Matrix inversion uses wavelength features and single-parameter calibration; fusion uses joint-parameter training and can also use power features. Supplementary Note~S3 compares wavelength regression, bias correction, and the addition of power predictions to show how each affects estimation error.

\subsection{Paired comparison and decomposition of systematic errors}

The population standard deviation of the errors, $s=\sqrt{N^{-1}\sum_i(\epsilon_i-\mathrm{MBE})^2}$, satisfies $\mathrm{RMSE}^2=\mathrm{MBE}^2+s^2$. The ratio $\mathrm{MBE}^2/\mathrm{RMSE}^2$ therefore gives the fraction of the mean-square error caused by a common offset, which reflects trueness rather than precision \cite{jcgm2012vim}. Because all methods are applied to the same spectra, we compare them by paired bootstrap resampling. Each of 10{,}000 resamples draws 18 spectra with replacement and uses the same draw for every method; the 2.5th and 97.5th percentiles of the resampled RMSE differences give a 95\% interval. We also count the spectra for which one method gives the smaller absolute error. These intervals describe variation among the 18 evaluation spectra only.

To trace the systematic errors, we fit a second-order forward model to each trough wavelength and minimum power using only the 27 training spectra:
\begin{equation}
\widehat x_k(\mathbf p)=a_{k0}+\sum_{m=1}^{3}\left(a_{km}p_m+q_{km}p_m^2\right),
\label{eq:cov_forward}
\end{equation}
where $p_1$, $p_2$, and $p_3$ denote $T$, $c$, and $h$. Leave-one-out residuals on the training spectra assess this model, and forward models with linear terms only or with additional interaction terms provide comparisons. The compared estimators are affine in the extracted features. For estimator $\widehat{\mathbf p}(\cdot)$, the error for evaluation spectrum $i$ with features $\mathbf x_i$ and reference values $\mathbf p_i$ therefore splits exactly as
\begin{equation}
\boldsymbol\epsilon_i=
\underbrace{\widehat{\mathbf p}\bigl(\widehat{\mathbf x}(\mathbf p_i)\bigr)-\mathbf p_i}_{\text{curvature component}}
+\underbrace{\widehat{\mathbf p}(\mathbf x_i)-\widehat{\mathbf p}\bigl(\widehat{\mathbf x}(\mathbf p_i)\bigr)}_{\text{spectral-difference component}}.
\label{eq:cov_decomposition}
\end{equation}
The first component is the error the estimator would make if the features followed the second-order training response exactly. For the regression estimators, this component would vanish, apart from negligible ridge shrinkage, if the response were linear and additive. It therefore measures the effect of response curvature that the linear estimators do not represent. For matrix inversion, it also includes differences between the single-parameter calibration and the joint training response. The second component propagates the difference between the measured features and the forward model. The evaluation reference values enter only this analysis, after all estimators were fixed.

\section{Results}
\label{sec:results}
\subsection{Independent calibration provides distinct wavelength responses}

\begin{figure*}[t]
\centering
\includegraphics[width=\textwidth]{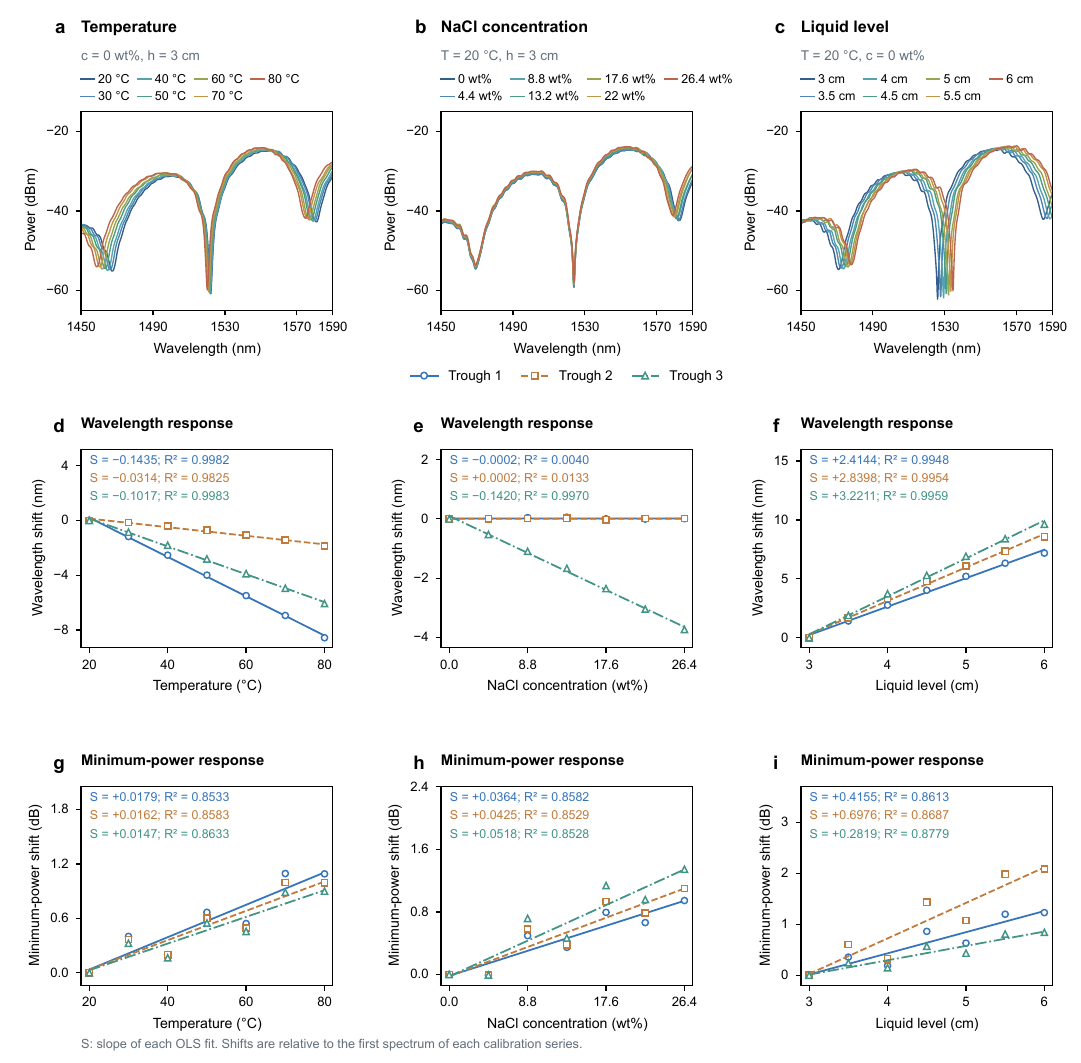}
\caption{Measured spectra and single-parameter calibration responses. Columns show temperature, NaCl mass fraction, and liquid level. (a)--(c) Seven measured spectra per parameter, displayed over 1450--1590~nm without smoothing or vertical offsets. (d)--(f) Three-trough wavelength shifts; (g)--(i) sampled-minimum-power changes in dB. Shifts and changes are relative to the first scan of each calibration series. Symbols show extracted features and lines show ordinary-least-squares fits. $S$ denotes the fitted slope, with wavelength units of nm/$^\circ$C, nm/pp, and nm/cm and power units of dB/$^\circ$C, dB/pp, and dB/cm for the respective columns; $R^2$ is the coefficient of determination. Temperature calibration holds $c=0$~wt\% and $h=3$~cm; concentration calibration holds $T=20~^\circ$C and $h=3$~cm; liquid-level calibration holds $T=20~^\circ$C and $c=0$~wt\%. Features and fits use the full 1400--1674~nm acquisitions and the stated extraction windows; the display crop omits trough-3 minima beyond 1590~nm at higher liquid levels.}
\label{fig:cov_calibration}
\end{figure*}

The seven spectra from each single-parameter calibration are shown over 1450--1590~nm in Fig.~\ref{fig:cov_calibration}(a)--(c). The corresponding wavelength shifts and minimum-power changes in Fig.~\ref{fig:cov_calibration}(d)--(i) are relative to the first scan of each calibration series. The display range does not restrict feature extraction or calibration fits, which use the full acquired spectra.

The three trough wavelengths responded differently to temperature in the single-parameter calibration (Fig.~\ref{fig:cov_calibration}(d)). Ordinary-least-squares fits gave slopes of $-0.1435$, $-0.0314$, and $-0.1017$~nm/$^\circ$C, with $R^2$ values of 0.99822, 0.98245, and 0.99831. Increasing liquid level shifted all three troughs toward longer wavelengths, with slopes of 2.4144, 2.8398, and 3.2211~nm/cm. Concentration mainly affected trough 3, whose wavelength slope was $-0.14195$~nm/pp with $R^2=0.99696$. Troughs 1 and 2 had small slopes and low $R^2$ values, indicating weak wavelength responses to concentration over the measured range.

The fitted wavelength matrix was full rank, with a column-normalized condition number of 4.383. The sampled-minimum-power calibration fits had $R^2$ values of 0.853--0.878 (Fig.~\ref{fig:cov_calibration}(g)--(i)). For each parameter, we calculated the wavelength and power models' prediction errors on held-out training folds and used these errors to determine their fusion weights. Supplementary Note~S1 reports all wavelength and power calibration fits.

\subsection{Errors of matrix inversion, wavelength regression, and fusion}

We compare matrix inversion, bias-corrected wavelength regression, and wavelength--power fusion on the same 18 evaluation spectra. The wavelength-only comparison uses the same fitted wavelength model and bias correction as fusion, with the power contribution omitted.

Relative to matrix inversion, fusion reduced liquid-level RMSE by 43.23\%, from 0.2448 to 0.1390~cm (Table~\ref{tab:cov_performance}). Fusion gave the smaller absolute error for all 18 spectra, and the paired 95\% bootstrap interval of the RMSE reduction was 0.1018--0.1104~cm (Table~\ref{tab:cov_paired}). Concentration RMSE decreased by 10.69\%, from 1.0594 to 0.9461~pp, with smaller errors for 16 of 18 spectra and an interval of 0.059--0.167~pp. Temperature RMSE was 0.4035~$^\circ$C for fusion and 0.4284~$^\circ$C for matrix inversion. This 5.82\% difference lay within its interval of $-$0.0496 to 0.1124~$^\circ$C, and fusion gave the smaller error for 10 of 18 spectra; the 18 spectra therefore do not resolve a temperature difference between the two methods. Before power estimates were added, bias-corrected wavelength regression gave liquid-level, concentration, and temperature RMSEs of 0.1478~cm, 0.9461~pp, and 0.4112~$^\circ$C, respectively.

\begin{table*}[t]
\centering
\caption{Errors of matrix inversion, bias-corrected wavelength regression, and wavelength--power fusion on the same 18 evaluation spectra. The wavelength-regression comparison uses the same fitted wavelength model and bias correction as fusion, with no power contribution. RMSE and MAE are reported for liquid level, concentration, and temperature. Concentration errors are expressed in mass-fraction percentage points.}
\label{tab:cov_performance}
\setlength{\tabcolsep}{7pt}
\begin{tabular}{lrrrrrr}
\toprule
Method & $h$ RMSE (cm) & $c$ RMSE (pp) & $T$ RMSE (${}^\circ$C) & $h$ MAE (cm) & $c$ MAE (pp) & $T$ MAE (${}^\circ$C) \\
\midrule
Independent matrix inversion & 0.2448 & 1.0594 & 0.4284 & 0.2412 & 0.9745 & 0.3522 \\
Bias-corrected wavelength & 0.1478 & 0.9461 & 0.4112 & 0.1413 & 0.8662 & 0.3291 \\
Wavelength--power fusion & 0.1390 & 0.9461 & 0.4035 & 0.1316 & 0.8662 & 0.3162 \\
\bottomrule
\end{tabular}
\end{table*}

\begin{table}[t]
\centering
\caption{Paired comparison of RMSE on the same 18 evaluation spectra. M, W, and F denote matrix inversion, bias-corrected wavelength regression, and wavelength--power fusion. $\Delta$RMSE is the RMSE of the first method minus that of the second, so positive values favor the second method. Intervals are percentile 95\% intervals from 10{,}000 paired bootstrap resamples. ``Improved'' counts the spectra for which the second method gives the smaller absolute error.}
\label{tab:cov_paired}
\small\setlength{\tabcolsep}{3pt}
\begin{tabular}{llrrr}
\toprule
Output & Comparison & $\Delta$RMSE & 95\% interval & Improved\\
\midrule
$h$ (cm) & M $\rightarrow$ W & +0.0970 & [+0.0945, +0.0998] & 18/18\\
 & W $\rightarrow$ F & +0.0088 & [+0.0043, +0.0137] & 15/18\\
 & M $\rightarrow$ F & +0.1058 & [+0.1018, +0.1104] & 18/18\\
\addlinespace
$c$ (pp) & M $\rightarrow$ W & +0.113 & [+0.059, +0.167] & 16/18\\
 & W $\rightarrow$ F & \multicolumn{3}{c}{identical (zero power weight)}\\
 & M $\rightarrow$ F & +0.113 & [+0.059, +0.167] & 16/18\\
\addlinespace
$T$ (${}^\circ$C) & M $\rightarrow$ W & +0.0172 & [$-$0.0494, +0.0882] & 11/18\\
 & W $\rightarrow$ F & +0.0077 & [$-$0.0287, +0.0567] & 7/18\\
 & M $\rightarrow$ F & +0.0249 & [$-$0.0496, +0.1124] & 10/18\\
\bottomrule
\end{tabular}
\end{table}

\begin{figure*}[t]
\centering
\includegraphics[width=\textwidth]{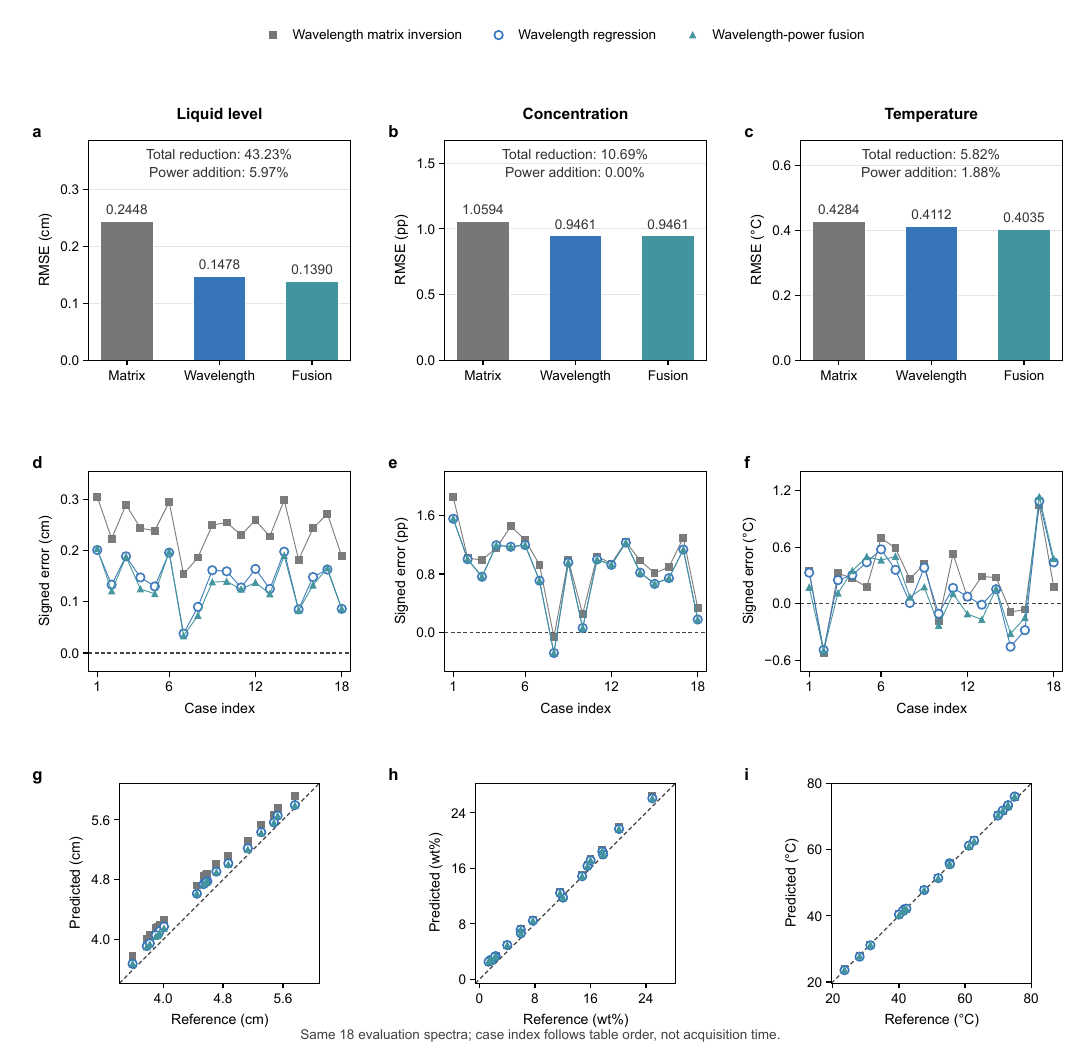}
\caption{Comparison of matrix inversion, bias-corrected wavelength regression, and wavelength--power fusion on the same 18 evaluation spectra. Columns show liquid level, NaCl concentration, and temperature. (a)--(c) RMSE across all 18 spectra; (d)--(f) signed prediction errors for each evaluation case; (g)--(i) predicted values against reference values, with dashed identity lines. Wavelength regression uses the same fitted wavelength model and bias correction as fusion, with no power contribution. Concentration predictions from wavelength regression and fusion coincide because the power weight is zero. Total reduction compares fusion with matrix inversion; power addition compares fusion with bias-corrected wavelength regression. Table~\ref{tab:cov_paired} gives paired bootstrap intervals for these differences. Case numbers follow the order in Supplementary Note~S2 and do not represent acquisition time.}
\label{fig:cov_evaluation}
\end{figure*}

\subsection{Wavelength and power contributions to the estimates}

The wavelength-model weights were 98.11\%, 100.00\%, and 96.90\% for temperature, concentration, and liquid level. The corresponding power-model weights were 1.89\%, 0.00\%, and 3.10\% (Table~\ref{tab:cov_parameters}). Each weight multiplies the corresponding bias-corrected model estimate and remains unchanged for all evaluation spectra. For concentration, the unconstrained wavelength weight was 1.1152. Applying the $[0,1]$ constraint set the wavelength weight to 1 and the power weight to 0. Under the fixed fusion settings, the training-error criterion therefore favored the bias-corrected wavelength estimate alone for concentration.

\begin{table}[t]
\centering
\caption{Fixed settings and model weights used to combine wavelength and power estimates. $\gamma$ sets the fraction of mean error subtracted from each estimate. $\rho$ scales the cross-error terms, and $\eta$ adjusts error terms involving the power model.}
\label{tab:cov_parameters}
\setlength{\tabcolsep}{4pt}
\begin{tabular}{lrrrrr}
\toprule
Output & $\gamma$ & $\rho$ & $\eta$ & $w_\lambda$ (\%) & $w_P$ (\%) \\
\midrule
$T$ & 0.5 & 0.0 & 1.0 & 98.1148 & 1.8852 \\
$c$ & 1.0 & 1.0 & 4.0 & 100.0000 & 0.0000 \\
$h$ & 0.0 & 0.0 & 4.0 & 96.9040 & 3.0960 \\
\bottomrule
\end{tabular}
\end{table}

Before bias correction, the wavelength model gave temperature, concentration, and liquid-level RMSEs of 0.4178~$^\circ$C, 0.9627~pp, and 0.1478~cm. After bias correction, these values were 0.4112~$^\circ$C, 0.9461~pp, and 0.1478~cm (Supplementary Note~S3). Adding the weighted power estimate further reduced liquid-level RMSE by 0.0088~cm, with an interval of 0.0043--0.0137~cm and smaller errors for 15 of 18 spectra. The corresponding temperature reduction of 0.0077~$^\circ$C lay within its interval of $-$0.0287 to 0.0567~$^\circ$C. The concentration power weight was zero, so the final concentration estimate came entirely from the bias-corrected wavelength model.

The jointly trained wavelength model reduced estimation errors before power predictions were added. Its liquid-level and concentration reductions relative to matrix inversion were 0.0970~cm and 0.113~pp, with intervals that excluded zero (Table~\ref{tab:cov_paired}). The overall reduction relative to matrix inversion therefore came mainly from the jointly calibrated wavelength model, with a smaller contribution from power for liquid level.

\begin{figure*}[t]
\centering
\includegraphics[width=\textwidth]{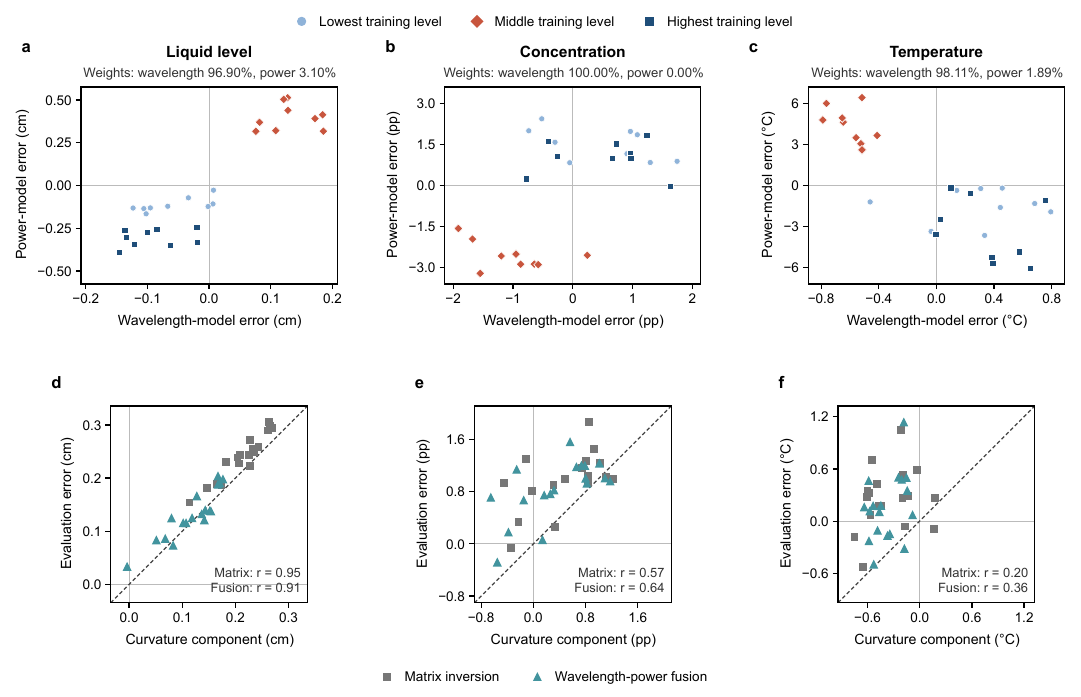}
\caption{Cross-fitted training errors, fixed weights, and the curvature component of the evaluation errors. (a)--(c) Wavelength-model and power-model errors on held-out training folds after bias correction, for the 27 training spectra. Colors mark the lowest, middle, and highest training level of the parameter in each column, and the headings give the fixed model weights. (d)--(f) Evaluation errors of matrix inversion and fusion against their curvature components, which are obtained by applying each estimator to features predicted by a second-order forward model fitted to the training spectra. Dashed lines show equality, and $r$ is the Pearson correlation across the 18 evaluation spectra. Panels (a)--(c) do not use the evaluation spectra.}
\label{fig:cov_weights}
\end{figure*}

\subsection{Cross-validation results}

Four tests used the training data: fivefold cross-validation and three tests that each omitted one parameter's middle level from model fitting (Supplementary Note~S2). The ratio of fusion RMSE to matrix-inversion RMSE was 0.722--0.846 for temperature, 1.064--2.689 for concentration, and 0.671--1.885 for liquid level. Fusion gave lower temperature RMSE but higher concentration RMSE in all four tests. Liquid-level RMSE was also higher when the middle liquid level was omitted from training. These results came from withheld training spectra, not from the 18 evaluation spectra. Section~\ref{sec:cov_curvature} relates the larger errors at withheld middle levels to curvature in the spectral responses.

\subsection{Systematic errors and response curvature}
\label{sec:cov_curvature}

The liquid-level and concentration errors were dominated by a common positive offset (Table~\ref{tab:cov_decomposition}). All 18 liquid-level errors were positive for every method. For fusion, the mean bias error (MBE) was +0.132~cm, whereas the scatter about it was 0.045~cm, so the offset accounted for 90\% of the mean-square error. Matrix inversion showed the same pattern, with an MBE of +0.241~cm and a scatter of 0.042~cm. For concentration, 17 of 18 errors were positive, and the offset accounted for 78--83\% of the mean-square error. Temperature errors were dominated by scatter, with offsets accounting for 14--36\%. The scatter differed little among the methods, so the liquid-level and concentration improvements came from smaller offsets.

\begin{table*}[t]
\centering
\caption{Decomposition of the evaluation errors on the same 18 spectra. MBE is the mean signed error and SD the population standard deviation about it, so that $\mathrm{RMSE}^2=\mathrm{MBE}^2+\mathrm{SD}^2$. MBE$^2$/MSE is the fraction of the mean-square error caused by the common offset, and $\epsilon>0$ counts positive errors. The last two columns split MBE into the mean curvature and spectral-difference components of (\ref{eq:cov_decomposition}). Units follow the output column.}
\label{tab:cov_decomposition}
\setlength{\tabcolsep}{5pt}
\begin{tabular}{llrrrrrrr}
\toprule
Output & Method & MBE & SD & RMSE & MBE$^2$/MSE & $\epsilon>0$ & Curvature & Spectral diff.\\
\midrule
$h$ (cm) & Matrix inversion & +0.241 & 0.042 & 0.245 & 97\% & 18/18 & +0.214 & +0.027\\
 & Bias-corrected wavelength & +0.141 & 0.043 & 0.148 & 91\% & 18/18 & +0.115 & +0.027\\
 & Wavelength--power fusion & +0.132 & 0.045 & 0.139 & 90\% & 18/18 & +0.119 & +0.013\\
\addlinespace
$c$ (pp) & Matrix inversion & +0.97 & 0.43 & 1.06 & 83\% & 17/18 & +0.49 & +0.48\\
 & Bias-corrected wavelength & +0.83 & 0.45 & 0.95 & 78\% & 17/18 & +0.36 & +0.47\\
 & Wavelength--power fusion & +0.83 & 0.45 & 0.95 & 78\% & 17/18 & +0.36 & +0.47\\
\addlinespace
$T$ (${}^\circ$C) & Matrix inversion & +0.257 & 0.343 & 0.428 & 36\% & 14/18 & $-$0.351 & +0.608\\
 & Bias-corrected wavelength & +0.180 & 0.370 & 0.411 & 19\% & 13/18 & $-$0.428 & +0.608\\
 & Wavelength--power fusion & +0.153 & 0.373 & 0.403 & 14\% & 12/18 & $-$0.374 & +0.527\\
\bottomrule
\end{tabular}
\end{table*}

The training errors did not show this offset because they varied with grid level. The cross-fitted liquid-level errors of the bias-corrected wavelength model averaged $-$0.057, +0.131, and $-$0.092~cm at 3, 4.5, and 6~cm, respectively (Fig.~\ref{fig:cov_weights}(a)). Their mean over the grid was close to zero, so the bias correction estimated from them could not remove positive errors at intermediate levels. Concentration and temperature errors showed the opposite pattern, with negative mean errors at the middle level (Fig.~\ref{fig:cov_weights}(b), (c)). Matrix inversion applied to the training spectra gave a liquid-level MBE of +0.099~cm, with positive errors for 25 of 27 spectra and the largest mean error, +0.224~cm, at the middle level. All evaluation liquid levels lay between 3.59 and 5.76~cm, inside the range where these level-dependent errors were positive.

These level-dependent errors reflect curvature in the wavelength responses. In the single-parameter calibrations, linear-fit residuals reached 0.32~nm for liquid level and 0.20~nm for temperature, about 9 and 6 times the largest repeated-scan standard deviation of 0.035~nm. These residuals were positive at intermediate settings and negative at both ends, indicating concave responses. Quadratic fits reduced the RMS residuals to at most 0.036 and 0.037~nm, respectively (Supplementary Note~S5). For the 27 training spectra, a forward model with linear terms only left leave-one-out wavelength residuals of 0.299--0.318~nm RMS. The second-order model in (\ref{eq:cov_forward}) reduced them to 0.027--0.045~nm, comparable to the repeated-scan scatter, and adding interaction terms did not reduce them further.

Applying each estimator to features predicted by this forward model reproduced most of the liquid-level offset (Table~\ref{tab:cov_decomposition} and Fig.~\ref{fig:cov_weights}(d)). The curvature component averaged +0.214~cm for matrix inversion and +0.119~cm for fusion, and it correlated with the individual errors ($r=0.95$ and 0.91, respectively). The spectral-difference component contributed +0.027 and +0.013~cm. Joint calibration therefore reduced the liquid-level offset mainly by reducing the curvature error: its regression was fitted across the joint grid, which centered the curvature errors over the calibrated range, whereas matrix inversion is anchored to the reference spectrum. The larger cross-validation errors when a middle level was withheld are consistent with this curvature, because a linear model fitted to the two outer levels cannot represent the response between them.

For concentration, the curvature and spectral-difference components contributed similarly to the fused offset (+0.36 and +0.47~pp). For temperature, the curvature component was negative ($-$0.374~$^\circ$C) and the spectral-difference component positive (+0.527~$^\circ$C). Relative to the forward model, the evaluation spectra had mean wavelength differences of $-$0.022, +0.058, and $-$0.042~nm for troughs 1--3, and trough 2 lay above the forward model for 17 of 18 spectra. Because temperature and concentration produce small wavelength shifts, the trough-2 difference alone changed the fused temperature and concentration estimates by about +0.411~$^\circ$C and +0.26~pp, but the liquid-level estimate by only +0.024~cm.

\subsection{Repeatability and stability under fixed conditions}

At three fixed conditions, we acquired 10 repeated scans per condition, giving 30 spectra in total. For each series, we calculated the sample standard deviations of the wavelength and minimum transmitted power of each of the three troughs. Across the series and troughs, wavelength standard deviations ranged from 0.0213 to 0.0350~nm, and minimum-power standard deviations ranged from 0.0064 to 0.0124~dB. These values quantify variation in the extracted spectral features during repeated acquisition at unchanged settings.

At each of three fixed conditions, stability measurements lasted 120~min, with one spectrum acquired every 4~min. Each series contained 31 spectra, including the initial measurement, giving 93 spectra in total. Across the series and troughs, wavelength standard deviations ranged from 0.0201 to 0.0387~nm, and minimum-power standard deviations ranged from 0.0347 to 0.0409~dB.

Wavelength standard deviations were of a similar magnitude in the two measurement types, whereas minimum-power standard deviations were higher in the stability series than in the repeated scans. In these measurements, the 120-min observation recorded greater power variation than repeated scanning. Power variation during observation should therefore be considered when using power information to improve estimation accuracy. Supplementary Note~S4 reports the statistics for each series and trough.

\section{Discussion}
\label{sec:discussion}
Our earlier work demonstrated multi-trough responses and dual-parameter discrimination \cite{xu2026multitrough}. Here, we extend this approach to simultaneous estimation under joint temperature, NaCl-concentration, and liquid-level variation. Established covariance-weighting methods provide a basis for combining wavelength and minimum-power estimates \cite{bates1969combination,bakr2018covariance}. Paired cross-fitted training errors determine regularized weights separately for each parameter, linking each readout's predictive reliability to its contribution to the final estimate. The matrix uses sensitivities measured by varying one parameter at a time, while the regression models use spectra measured at combinations of all three parameters.

Fusion reduced liquid-level and concentration RMSE by 43.23\% and 10.69\%, respectively, compared with matrix inversion, and the paired bootstrap intervals of both reductions excluded zero. The 5.82\% temperature difference was not resolved by the 18 spectra. Most of the improvement came from fitting the wavelength model to joint calibration data. Adding power estimates to the bias-corrected wavelength estimates reduced liquid-level RMSE by a further 0.0088~cm, with smaller errors for 15 of 18 spectra. This contribution was small but resolved. The temperature change from adding power was within its interval, and power did not contribute to concentration because its weight was zero. The weighting therefore gave the less accurate power estimate small weights for liquid level and temperature and excluded it for concentration, where the training-error criterion favored the wavelength estimate alone. The overall improvement over matrix inversion cannot therefore be attributed mainly to adding power features. Supplementary Note~S3 compares the wavelength model before and after bias correction with the final fused estimates.

The error decomposition identifies response curvature as the main remaining limitation. Both the single-parameter calibrations and the joint training spectra showed concave wavelength responses to liquid level and temperature, with linear-fit residuals up to about nine times the repeated-scan scatter. A linear estimator fitted over a three-level grid leaves training errors that average close to zero but are positive at the middle level and negative at the ends. A bias correction estimated from these errors is therefore small and cannot remove the positive offset at intermediate conditions, where all 18 evaluation spectra lay. Matrix inversion showed the same offset even though its column-normalized condition number was only 4.383. The condition number describes how random feature errors are amplified, but not systematic errors caused by an inadequate response model. Nonlinear wavelength responses have similarly been reported as a source of sensitivity-matrix errors in other fiber sensors \cite{chen2023seawaterml,wei2025cnnbilstm}.

Second-order calibration terms address this source directly. In repeated fivefold cross-validation on the 27 training spectra, adding squared wavelength terms to the wavelength regression reduced liquid-level RMSE from 0.110 to 0.066~cm, while temperature and concentration RMSEs changed from 0.512 to 0.530~$^\circ$C and from 1.05 to 1.07~pp (Supplementary Note~S5). The fusion rule does not require linear models, so curvature-aware wavelength and power models can be combined using the same error-based weights. Each parameter needs at least three calibration levels to identify a quadratic term, and additional intermediate levels would make the curvature estimate less dependent on the middle level. Because the present evaluation spectra were used to diagnose the curvature, such models should be assessed on new measurements.

The probe was remounted between the joint training and evaluation measurements. Relative to the second-order training response, the evaluation spectra showed mean wavelength differences of $-0.022$ to $+0.058$~nm, comparable to the 0.020--0.039-nm wavelength standard deviations observed over 120~min at fixed conditions. These differences changed the fused liquid-level estimate by only 0.013~cm on average. They shifted the temperature and concentration estimates more because these parameters produce small wavelength shifts. They may reflect remounting, solution preparation, or drift between the measurement sets, which the present data cannot distinguish. Measurements under simultaneous parameter changes have also given larger errors than test sets used to evaluate a trained model in another fiber sensor \cite{lin2024tfbgsagnac}. A spectrum acquired at a known reference condition after each installation, together with repeated remounting tests, would quantify this contribution.

Omitting a complete parameter level from training tested predictions at combinations that were absent from the model fit. The 18 evaluation spectra provided a separate comparison using measurements collected during early exploration. If those spectra informed earlier method choices, this comparison cannot be treated as a fully independent test, and the bootstrap intervals do not account for that possibility. The cross-validation results showed that fusion did not improve every parameter under every training design. Its larger maximum temperature error also showed that lower RMSE did not guarantee smaller errors for every measurement.

The model estimates each spectrum separately, with no temporal model or automatic update of calibration coefficients or weights. Applying it to another device, liquid, or measurement session therefore requires checking whether the calibration still describes the measured responses. Measurements with different devices and on different days would test whether the fitted models and weights remain useful without recalibration. Other intensity features could also be tested to determine whether they improve predictions beyond those obtained from trough wavelengths alone.

\section{Conclusion}
We compared wavelength-matrix inversion, bias-corrected wavelength regression, and covariance-weighted wavelength--power fusion for estimating temperature, NaCl concentration, and liquid level from measured fiber-sensor spectra. Single-parameter calibration gave a full-rank wavelength matrix with a column-normalized condition number of 4.383. Paired cross-fitted errors from 27 joint calibration conditions determined bias corrections and fixed weights separately for each parameter, linking the contribution of each readout to its predictive reliability. The power weights were 1.89\%, 0.00\%, and 3.10\% for temperature, concentration, and liquid level.

On the 18 evaluation spectra, fusion reduced liquid-level RMSE from 0.2448 to 0.1390~cm and concentration RMSE from 1.0594 to 0.9461~pp, and paired bootstrap intervals excluded zero for both reductions. Temperature RMSEs of 0.4035 and 0.4284~$^\circ$C were not resolved. Most of the improvement came from the jointly calibrated wavelength model; relative to it, power further reduced liquid-level RMSE by 0.0088~cm, while the concentration power weight was zero. The remaining liquid-level and concentration errors were dominated by positive offsets. A second-order forward model fitted to the training spectra attributed most of the liquid-level offset to curvature in the trough-wavelength responses, which linear estimators and training-error bias corrections cannot remove at intermediate conditions. Small spectral differences between the training and evaluation measurements contributed most of the temperature offset. These results support evaluating auxiliary spectral features by the information they add to parameter estimates, including features with weaker single-parameter calibration fits, and they identify curvature-aware calibration as the next step toward higher accuracy.

\clearpage
\IEEEtriggeratref{48}
\bibliographystyle{IEEEtran}
\bibliography{refs}

\end{document}